\documentclass[12pt]{article}
\usepackage{amsmath,amssymb,epsfig,cite,color}
\advance\hoffset by -1.1truecm
\advance\voffset by -1.0truecm
\newcommand{\be}{\begin{equation}}
\newcommand{\ee}{\end{equation}}
\newcommand{\bea}{\begin{eqnarray}}
\newcommand{\eea}{\end{eqnarray}}

\usepackage{authblk}
\usepackage{graphicx}

\newcounter{appendice}

\numberwithin{equation}{section}

\begin{document}

\title{
\begin{flushright}
\small
\end{flushright}
\vspace{2.2cm}\begin{flushleft} \bf{Localisation in Quantum Field Theory\footnote{
This article is a brief review of work done  by Brunetti, Guido and Longo \cite{HaagBook,BGLS}, Schroer  and Fassarella and Schroer \cite{SF},  
Mund, Schroer and Yngvaso \cite{large} and others on localisation  problems in relativistic quantum field theories and their deep implications. The review is informal in that mathematical rigour is not attempted. It is at a level accessible to most quantum field theorists. 
It is dedicated to Peter Presnajder, wonderful friend and close collaborator.}
}\end{flushleft}}
\author{A. P. Balachandran}
\affil{Physics Department, Syracuse University, Syracuse, New York 13244-1130, U.S.A \\ 
Institute of Mathematical Sciences, C.I.T Campus, Chennai, TN 600113, India}

\date{\empty}

\maketitle
\begin{abstract}
In nonrelatistic quantum mechanics, Born's principle of localistion is as follows: For a single particle, if a wave function $\psi_K$ vanishes outside a spatial region $K$, it is said to be localised in $K$. In particular if a spatial region $K'$ is disjoint from $K$, a wave function $\psi_{K'}$ localised in $K'$ is orthogonal to $\psi_K$.

Such a principle of localisation does not exist compatibly with relativity and causality in quantum field theory (Newton and Wigner) or interacting point particles (Currie,Jordan and Sudarshan).It is replaced by symplectic localisation of observables as shown by Brunetti, Guido and Longo, Schroer and others. This localisation gives a simple derivation of the spin-statistics theorem and the Unruh effect, and shows how to construct quantum fields for anyons and for massless particles with `continuous' spin.

This review outlines the basic principles underlying symplectic localisation and shows or mentions its deep implications. In particular, it has the potential to affect relativistic quantum information theory and black hole physics.
\end{abstract}
\section{Introduction}


Locality in quantum field theory is often used in the sense that test functions had support in a localised region of a spatial slice or in spacetime. 
This interpretation is suggested by Born's interpretation of wave functions  psi  of a particle :  $|\psi(x)|^2 d^3x$ is the probabilty of finding the particle in a voume $d^3x$ around $x$.

While this interpretation is adequate for a first approach, it becomes incomplete when the requirements of
relativistic invariance and causality are brought in. A more sophisticated aproach becomes necessary.

In theories with no gauge invariance, the full set of axioms for local relativistic quantum physics
has been developed by Haag and Kastler and discussed in Haag's book~\cite{HaagBook}.
The notes that follow will not discuss the Haag-Kastler approach, but will borrow ideas therefrom to describe this more refined approach..

As we explain below, the idea of localisation of wave functions requires the existence of a position operator.
That is problematic in relativistic quantum physics.
Instead, for relativistic free fields, a new concept of localisation, which localises observables instead of states,
has been formulated by Brunetti, Guido and Longo \cite{BGLS} and by Schroer and colleagues~\cite{SF,large}. 
It gives new insights about particles obeying braid statistics, and those transforming by massless
``continuous spin'' representations.


\section{On Position Operators}

In quantum physics, just as in classical physics, observables $\mathcal{A}$ determine the measurements available on the system.
They form an algebra. That is, if $\alpha,\beta$ are observables, we have a multiplication map $m$ from
$\mathcal{A}\otimes\mathcal{A}$ to $\mathcal{A}$:
\begin{eqnarray}
m:\,\, \alpha\otimes\beta \,\,\rightarrow \,\, m(\alpha\otimes\beta ):=\alpha\beta,
\end{eqnarray}
which is linear in each entry. The algebra $\mathcal{A}$ in quantum physics is non-commutative, whereas 
the corresponding algebra $\mathcal{A}_c$ is commutative in classical physics.

The classical algebra $\mathcal{A}_c$ can be realised as real functions $f$ on the phase space
$T^\ast Q$ with (local) coordinates $(q_1,\ldots,q_N;p_1,\ldots,p_N):=(q,p)$:
\begin{eqnarray}
\begin{array}{llll}
f: &\,\, T^\ast Q\,\,\rightarrow & \mathbb{R} \\ \\
& \overline{f(q,p)}=& f(q,p).
\end{array}
\end{eqnarray}
The property which corresponds to reality  in quantum physics is that there is a star operation or ``hermitean conjugation'' $\ast$
defined on $\mathcal{A}$:
\begin{eqnarray}
\alpha \in \mathcal{A} \quad \rightarrow \quad \alpha^\ast \in \mathcal{A}.
\end{eqnarray}

The outcome of experiments in classical physics is given by a probability distribution $\rho_c$ on $\mathcal{A}_c$.
It has the following basic properties:
\begin{itemize}
\item $\rho_c(q,p)\geq 0$.
\item $\int \, d\mu(q,p)\, \rho_c(q,p)=1$.
\item Mean value of $a_c\in\mathcal{A}_c:=\langle a_c\rangle=\int \,d\mu(q,p)\,\rho_c(q,p)\, a_c(q,p)\in\mathbb{R}$.
\end{itemize}
Here $d\mu$ is the Liouville volume form on $T^\ast Q$: $d\mu(q,p)=dq_1\wedge \ldots \wedge dq_N\wedge dp_1\wedge
\ldots \wedge dp_N$.

Correspondingly, in quantum physics, we have a {\it state} $\omega$ on $\mathcal{A}$.
It is a linear map with $\omega(a)$ giving the mean value of $a\in \mathcal{A}$. It has the properties:
\begin{itemize}
\item $\omega(a^\ast a)\geq 0$.
\item $\omega(\mathbb{I})=1$.
\item $\omega(a^\ast)=\overline{\omega(a)}$.
\end{itemize}
Here, the first two properties adapt all the properties of $\rho_c$ before, while the last property preserves the $\ast$
of $\mathcal{A}$ as complex conjugation on $\mathbb{C}$.

It is a theorem of Gelfan'd, Naimark and Segal (GNS) that, given a state $\omega$ on $\mathcal{A}$,
there exists a Hilbert space $\mathcal{H}$ on which 
$\mathcal{A}$ is realised as an algebra of operators, still denoted by $\mathcal{A}$ by us. 
Also, the $\ast$-operator becomes the hermitean adjoint $\dagger$.
Finally, $\omega$ can be represented by a density matrix $\rho$:
\begin{eqnarray}
\rho=\sum | \psi_i\rangle \langle \psi_i|, \quad\quad |\psi_i\rangle \in \mathcal{H}, \quad\quad \textrm{Tr}\rho=1.
\end{eqnarray}

From this abstract formulation, the wave function $\psi$ in non-relativistic quantum mechanics is recovered as follows.
The algebra $\mathcal{A}$ has an operator $\hat{x}$, called the position operator, with commuting components. 
If $|\vec{x}\rangle$ is the eigenstate of $\hat{x}$, 
\begin{eqnarray}
\hat{x}_i|\vec{x}\rangle= x_i |\vec{x}\rangle, \quad\quad \langle \vec{x}^\prime |\vec{x}\rangle=\delta^d(x^\prime-x),
\end{eqnarray}
where the spatial dimension $d$ is 3 for $\mathbb{R}^3$, we can write
\begin{eqnarray}
\psi(\vec{x})=\langle \vec{x}|\psi\rangle
\end{eqnarray}
for a vector $|\psi\rangle \in \mathcal{H}$ of norm 1: $\langle\psi|\psi\rangle=1$, associated with the rank 1 
(pure) state $\rho=|\psi\rangle\langle\psi|$. Then, $\psi$ gives the wave function of non-relativistic
physics, subject to Born's interpretation: $|\psi(\vec{x})|^2$ is the probability density for finding the system at
$\vec{x}$.

It is important that $\hat{x}$ transforms correctly under the Galilei group which is the governing group of
non-relativistic physics. Thus, it is a rotational vector and under a spatial translation $\vec{a}$, it changes to
$\hat{x}+\vec{a}$.

For a special relativistic particle, the Galilei group is changed to the Poincar\'{e} group $\mathcal{P}$, which has
the Lorentz group $\mathcal{L}$ as a subgroup.
The Lorentz group $\mathcal{L}$ transforms the spacetime point $x=(x_0,\vec{x})$ to $\Lambda x=(\Lambda_\mu^{{\color{white}
\mu}\nu} x_\nu)$. It transforms time and the new time $(\Lambda x)_0$ depends on the old spatial coordinate
$\vec{x}$. This fact leads to the disturbing result that a covariant position operator $\hat{x}=(\hat{x}^0,\hat{\vec{x}})$
does not exist for interacting relativistic particles. This basic result is due to Curry, Jordan and
Sudarshan~\cite{CJS}. The requirement of a covariant position operator is also called the world line condition and discussed
in Sudarshan and Mukunda~\cite{CJS}.

In relativistic quantum field theory, a similar situation prevails. The Newton-Wigner position operator~\cite{NewWig} is not
covariant and unsuitable for discussion of, say, causality.

The conclusion is that Bohr's interpretation of quantum mechanics cannot be adapted to relativistic systems.

But we need the notion of spacetime localisation.
It is a central element in formulating causality: this is the requirement that if spacetime regions $K_1$ and $K_2$
are spacelike separated, the corresponding observables commute.
It is also needed to interpret the statement that measurements are done on observables localised in a spacetime region $K$.
Such a notion, called ``modular localisation'', will be described below.

We now informally indicate the reason why the covariant position operator does not exist in a relativistic theory
in the presence of interactions.

\subsection{On Covariant Position Operators}

We illustrate the problem by considering the case of $N$ point particles with masses $m_i$. If
$Z(\tau^{(i)})=(Z_\mu (\tau^{(i)}))\in\mathbb{R}^4$ are the trajectories of the particles labelled by the parameters
$\tau^{(i)}\in(-\infty,\infty)$ and if the particles are non-interacting, we can describe them by the following
Lagrangian:
\begin{eqnarray}
\mathcal{L}=\sum_i\mathcal{L}^{(i)}, \quad\quad \mathcal{L}^{(i)}=m_i \sqrt{\left(\frac{dZ(\tau^{(i)})}{d\tau^{(i)}}\right)^2}.
\end{eqnarray}
The corresponding action is
\begin{eqnarray}
S=\sum_i S^{(i)}, \quad\quad S^{(i)}=\int d\tau^{(i)}\mathcal{L}^{(i)}.
\end{eqnarray}

This action is perfectly compatible with Poincar\'{e} invariance.
It can be quantised~\cite{GTandFB}.
Each $S^{(i)}$ gives the unitary irreducible representation (UIRR) of the Poincar\'{e} group for mass $m_i$ and spin 0. 
If $\mathcal{H}^{(i)}$ is the Hilbert space carrying the UIRR for the $i$-th particle, the full Hilbert
space is $\mathcal{H}=\otimes_i \mathcal{H}^{(i)}$. Thus we get the tensor product of $N$ UIRR's.

Now suppose that we wish to put in interactions.
They will couple different $Z(\tau^{(i)})$'s.
That will involve the identification of different $\tau^{(i)}$'s,
that is, effectively of different time coordinates, in some fashion.

But there is no consistent manner to do so since, as remarked above, Lorentz transformations change time in a manner
which involves spatial coordinates.

In the literature, there are many attempts to overcome this ``no-interaction theorem'', but none of them have led to
a satisfactory approach, compatible with causality and Poincar\'{e} invariance.

In a quantum field theory (QFT), the position operator has to be constructed using the quantum field $\varphi$,
so that it is covariant.
Early attempts to construct a position operator by Newton and Wigner~\cite{NewWig} and others did not succeed
in finding such a four-vector.

References from which literature after~\cite{CJS,NewWig} can be traced are~\cite{BalDom} and recent papers involving Schroer.

\subsection{The Two Concepts of Localisation}

Earlier, it was emphasised that both classical and quantum physics are formulated using
the concepts of both states and observables. Thus, we can study the localisation of either states or observables
(or perhaps both). 

In non-relativistic physics, it so happens that either localisation implies the other. We can informally
explain why that is so. If $K$ is a bounded spatial region and
\begin{eqnarray}
P_K=\int_K d^dx |\vec{x}\rangle \langle\vec{x}|
\end{eqnarray}
is the projection operator which projects vectors $\psi$ in the Hilbert space $\mathcal{H}$ to vectors
with support in $K$, 
\begin{eqnarray}
\psi_K=P_K|\psi\rangle=\int_K d^dx |\vec{x}\rangle \psi(\vec{x}),
\end{eqnarray}
then, for two such vectors $\psi_K,\chi_K$,
\begin{eqnarray}
\langle \chi_K| a| \psi_K\rangle=\int d^3x\, d^3y\, \langle \chi|P_KaP_K|\psi\rangle.
\end{eqnarray}
This shows that we can restrict wave functions to $K$, or equivalently restrict obsservables $a$ to $K$
by considering $P_KaP_K$.

But this reciprocity between localised states and localised observables fails in relativistic theories.
We cannot localise states as discussed above. But we {\it can} localise observables.
It is this localisation that we discuss below.

\section{Localisation in QFT}

\subsection{Preliminaries}

We consider only free fields. We also restrict attention for now to a relativistic free field $\varphi$ of spin zero so that
\begin{eqnarray}
\begin{array}{llll}
&\varphi(x)=\int\frac{d^3k}{2k_0(2\pi)^3}\left(a_ke^{-ik\cdot x}+a_k^\dagger e^{ik\cdot x}\right), \\ \\
&[a_k,a_{k^\prime}^\dagger]=2k_0(2\pi)^3\delta^3(k-k^\prime), \quad\quad k_0=\sqrt{|\vec{k}|^2+m^2}.
\end{array}
\end{eqnarray}
(The metric is $(1,-1,-1,-1)_{\textrm{diagonal}}$.)

The associated Hilbert space carries a (anti-)unitary irreducible representation $\rho$ of the Poincar\'{e} group
including total spacetime reflection  (which is here identified with CPT).

The commutator $D$ of $\varphi$ at $x$ and $y$ is the causal function $D$:
\begin{eqnarray}
&& [\varphi (x),\varphi(y)]=D(x-y), \label{comm} \\
&& D(x)=\int\frac{d^3k}{2k_0(2\pi)^3}\left(e^{-i k\cdot x}-e^{ik\cdot x}\right).
\label{9.2}
\end{eqnarray}

\subsection{The Weyl algebra $\mathcal{W}$}

Let $f$ be a real test function for $\varphi$. That means that if $f$ has support in a spacetime region $K$,
\begin{eqnarray}
\varphi(f)=\int d^4x\, f(x)\varphi(x) \label{9.1}
\end{eqnarray}
is the field $\varphi$ localised in $K$.

In the absence of a good notion yet of localisation, this remark needs clarification. It will emerge later.
For now, we use it to derive the Weyl algebra.

Following Weyl, we replace the unbounded operator $\varphi(f)$ by the unitary operator
\begin{eqnarray}
W(f)=e^{\frac{i}{\sqrt{2}}\varphi(f)}.
\end{eqnarray}
As a consequence of (\ref{9.2}), $W$'s fulfill
\begin{eqnarray}
W(f)W(g)=W(f+g)e^{-\frac{i}{4}\sigma(f,g)}, \quad\quad \sigma(f,g)=\textrm{Im}\int d^4x\, d^4y\, f(x)D(x-y)g(y).
\end{eqnarray}
Here, $\sigma(f,g)=-\sigma(g,f)$. Also, since 
\begin{eqnarray}
(\Box +m^2)D(x)=0,
\end{eqnarray}
we have
\begin{eqnarray}
\sigma \left( (\Box+m^2)\alpha,g\right)=0
\end{eqnarray}
for functions $\alpha$ of compact support (say).
Modulo such functions, $\sigma$ can be shown to be a symplectic form on test functions. 

Let us introduce a scalar product on $f$'s using Fourier transform:
\begin{eqnarray}
\tilde{f}(k)=\int d^4x\, f(x)\, e^{ik\cdot x}, \quad\quad k_0=\sqrt{\vec{k}^2+m^2}. \label{10.1} 
\end{eqnarray}
\begin{eqnarray}
(f,g)=\int \frac{d^3k}{2k_0(2\pi)^3}\bar{\tilde{f}}(k)\tilde{g}(k). \label{10.2}
\end{eqnarray}
Then, we can write
\begin{eqnarray}
W(f)W(g)=W(f+g)\, e^{-\frac{i}{2}\textrm{Im}(f,g)}. \label{10.3}
\end{eqnarray}
In addition, we have the $\ast$-relation
\begin{eqnarray}
W(f)^\ast =W(-f). \label{10.4}
\end{eqnarray}

Equations (\ref{10.3}) and (\ref{10.4}) are the defining relations for the Weyl algebra $\mathcal{W}$.
The quantisation of the free field can be recovered from the quantisation of $\mathcal{W}$.

\subsection{Quantisation of $\mathcal{W}$: the Fock Space}

We now specialise to a real scalar field so that, from the vacuum, it creates an irreducible representation space of
the Poincar\'{e} group.
So the field $\varphi$ in (\ref{comm}) is ``hermitean''.

The quantisation of $\mathcal{W}$ as it emerges from the Fock space quantisation of $\varphi$ is the following.
Let $\mathcal{H}$ be the Hilbert space with the scalar product introduced above. Then, the
bosonic Fock space $\mathcal{F}(\mathcal{H})$ is
\begin{eqnarray}
\textrm{exp}(\mathcal{H}):=\mathbb{C} e_0\oplus \mathcal{H} \oplus \mathcal{H} \otimes_S \mathcal{H}\oplus \ldots,
\label{11.1}
\end{eqnarray}
where $e_0$ is the vacuum state with norm 1 and $\oplus$ denotes symmetrised tensor product.
Then,
\begin{eqnarray}
W(f)e_0=e^{-\| f\|^2/4}\,e^{\frac{i}{\sqrt{2}}f}\in \mathcal{F}(\mathcal{H}). \label{11.2}
\end{eqnarray}

\subsection{An Abstract Definition of Weyl Algebra}

We can now state this result for a real scalar field in a more convenient and abstract manner.
Let $\mathcal{H}$ be a complex Hilbert
space.
Let $\textrm{Re}\, \mathcal{H}$ be a ``real'' subspace of $\mathcal{H}$ so that it is closed only for real
linear combination of its vectors. Then, consider operators $W(h)$ labelled by $h\in\textrm{Re}\, \mathcal{H}$
and fulfilling the algebraic relations
\begin{eqnarray}
W(h_1)W(h_2)=W(h_1+h_2)e^{-\frac{i}{2}\textrm{Im}(h_1,h_2)}, \quad\quad W^\ast(h)=W(-h). \label{12.1}
\end{eqnarray}
The algebra generated by the $W$'s is the Weyl algebra $\mathcal{W}(\textrm{Re}\, \mathcal{H})$.

We can find a representation of $\mathcal{W}_{\textrm{Re}\, \mathcal{H}}$ following (\ref{11.1}) and (\ref{11.2}).

\subsection{Remarks}

 The way we pick $\textrm{Re}\, \mathcal{H}$ in further developments is by constructing an anti-linear
involution $S$:
\begin{eqnarray}
 S^2=\mathbb{I}. \label{12.2}
\end{eqnarray}
Then,
\begin{eqnarray}
S\zeta=\zeta \quad \textrm{if } \zeta\in \textrm{Re}\, \mathcal{H}. \label{12.3}
\end{eqnarray}
 The subspace $\textrm{Re}\, \mathcal{H}$ is said to be ``standard'' if 
\begin{eqnarray}
\overline{\textrm{Re}\, \mathcal{H} \oplus i \textrm{Re}\, \mathcal{H}}=\mathcal{H}, \quad\quad \textrm{Re}\, \mathcal{H}\cap i
\textrm{Re}\, \mathcal{H}=\{0\}.
\end{eqnarray}
The bar means closure in the Hilbert space norm.

In this case we can unambiguously decompose a vector $\eta \in \mathcal{H}$ into its ``real'' and ``imaginary'' parts
$\textrm{Re}\, \eta$ and $\textrm{Im}\, \eta$:
\begin{eqnarray}
\begin{array}{lll}
&\textrm{Re}\,\eta:=\frac{1}{2}(\mathbb{I}+S)\eta, &\quad\quad \textrm{Im}\,\eta:=-\frac{i}{2}(\mathbb{I}-S)\eta, \\ \\
&\eta=\textrm{Re}\,\eta +i\textrm{Im}\,\eta, &\quad\quad S\eta=\textrm{Re}\,\eta-i\textrm{Im}\,\eta. \label{13.1}
\end{array}
\end{eqnarray}

If an anti-linear involution $S$ gives a ``standard'' decomposition of $\mathcal{H}$ into
$\textrm{Re}\,\mathcal{H} \oplus i \textrm{Re}\,\mathcal{H}$ on using (\ref{12.3}), $S$ is said to be 
the Tomita-Takesaki operator (in its real version).

\section{Quantum Field Theory: Requirements on Localisation}

As alluded to before, we will localise the algebra of observables, that is the Weyl algebra.
The localised algebras will be presented abstractly in terms of real subspaces defined using Tomita-Takesaki
involutions. 
Their interpretation as algebras localised in spacetime regions will subsequently emerge.

We will {\it not} try to localise states. We {\it cannot} do that.

Consider a spacetime region $K$. Then, let $K^\prime$ denote its causal complement, so that if
$x\in K$, $x^\prime\in K^\prime$, then $x$ and $x^\prime$ are spacelike separated.

The given region $K$ is said to be {\it causally complete} if $K^{\prime\prime}=K$.

We will index a family of Weyl algebras by causally complete $K$, writing $\mathcal{W}(K)$ for the indexed algebra.
But to physically interpret $\mathcal{W}(K)$ as the algebra of observables localised in $K$, it must have the following
properties, which are physically well motivated:
\begin{itemize}
\item {\it Covariance}:
Let $\mathcal{P}_+$ denote the Poincar\'{e} group including the total reflection $R$: $R\vartriangleright
(x_0,x_1,x_2,\ldots)=
(-x_0,-x_1,-x_2,\ldots)$. Let $g\in\mathcal{P}_+$. It acts on $K$: $K\rightarrow gK$. We require that there is a
representation $\rho$ of $\mathcal{P}_+$ on $\mathcal{H}$ where $\rho(g)$ is unitary if $g\in\mathcal{P}_+^\uparrow$
and anti-unitary if $g\in R\mathcal{P}_+^\uparrow$, such that
\begin{eqnarray}
\mathcal{W}(gK)=\rho(g)\mathcal{W}(K)\rho^{-1}(g). \label{15.1}
\end{eqnarray}
The operator $\rho(R)$ will be denoted by $\Theta$. It is anti-unitary.
\item Haag duality  which implies causality: Let $\mathcal{W}^\prime(K)$ denote the commutant of $\mathcal{W}(K)$. Then,
$\mathcal{W}(K^\prime)=\mathcal{W}^\prime(K)$.
\item Isotony: If $K_1\subseteq K_2$, then $\mathcal{W}(K_1)\subseteq \mathcal{W}(K_2)$.
\end{itemize}

\section{The Construction of $\mathcal{W}(K)$}

As stated above, we can assume that we are given a representation $\rho$ of $\mathcal{P}_+$ on $\mathcal{H}$.
We assume it to be (anti-)unitary, irreducible (UIRR) and of {\it positive energy}, $p_0>0$. For now, we consider
the spin zero representation.

The net of local algebras emerges just from the UIRR's of $\mathcal{P}_+$, that is from Wigner's original research.
It does not appeal to classical concepts like Lagrangians and actions. This is a remarkable fact.

Fix a wedge $W$, say
\begin{eqnarray}
W=\{x\in M^4: \,\, x_1>|x_0|\}. \label{16.1}
\end{eqnarray}
It is used as a device to label the Weyl algebras, even the existence of spacetime need not enter in its conception.
We classify them below.

Then, the Lorentz boosts
\begin{eqnarray}
\Lambda_W(t)=\left(
\begin{array}{cccccccc}
\cosh t & -\sinh t & 0 & 0 \\
-\sinh t & \cosh t & 0 & 0 \\
0 & 0 & 1 & 0 \\
0 & 0 & 0 & 1
\end{array}
\right)
\end{eqnarray}
leave $W$ invariant:
\begin{eqnarray}
\Lambda_W(t)W=W.
\end{eqnarray}
It is contained in the stability group of $W$.
The full stability group is generated by these Lorentz boosts and rotations and translations of the $x_2-x_3$ plane.

Consider the $x_0-x_1$ reflection $j_W$:
\begin{eqnarray}
j_W \vartriangleright(x_0,x_1,x_r)=(-x_0,-x_1,x_r), \label{17.2}
\end{eqnarray}
where $x_r$ denotes the remaining spatial coordinates. It maps $W$ to its causal complement $W^\prime$.
The figure 1 shows $W$, $W^\prime$ and $j_W$ for  $(1+1)$ spacetime.
\begin{figure}[h]
\begin{center}
\includegraphics[scale=0.5]{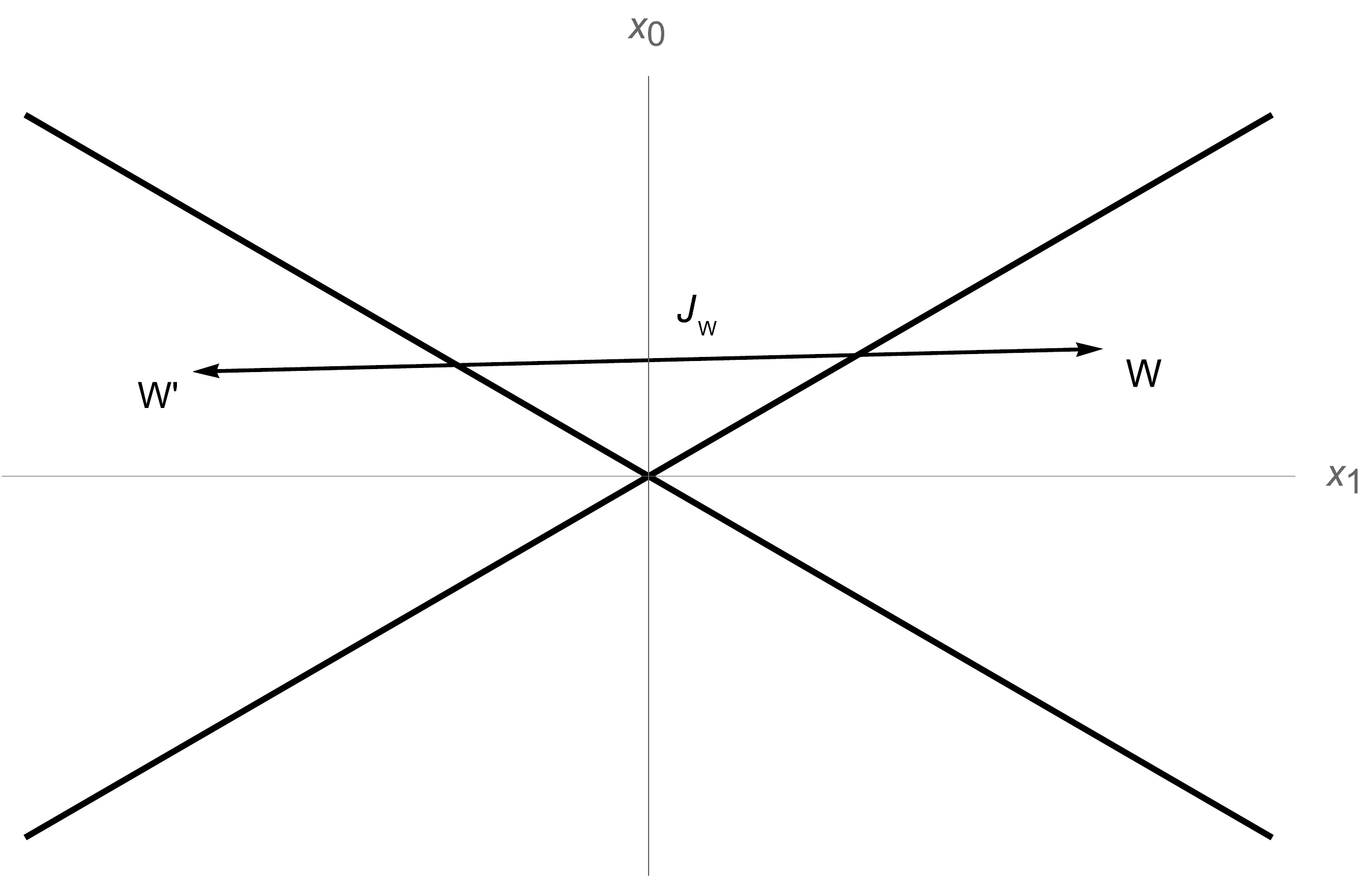}
\caption{ $W$, $W'$ and $j_W$  for (1 + 1)-dimensional spacetime.}
\end{center}
\end{figure}

An important property of $j_W$ is that it commutes with $\Lambda_W(t)$:
\begin{eqnarray}
j_W\Lambda_W(t)=\Lambda_W(t)j_W. \label{17.22}
\end{eqnarray}

Under the representation map $\rho$, $\Lambda_W(t)$ becomes
\begin{eqnarray}
\rho(\Lambda_W(t))=e^{itK_W}, \label{17.3}
\end{eqnarray}
while
\begin{eqnarray}
J_W:=\rho(j_W)=\Theta \times \rho( \pi-\textrm{rotation around 1-axis}). \label{18.1}
\end{eqnarray}
The expression for $J_W$ in terms of $\Theta$ assumes that the spacetime is four-dimensional and follows from
\begin{align}
\nonumber
(R\circ \pi-\textrm{rotation around 1-axis})\vartriangleright (x_0,x_1,x_r)=&R\vartriangleright(x_0,x_1,-x_r) 
=(-x_0,-x_1,x_r)\\=&j_W\vartriangleright(x_0,x_1,x_r). \label{19}
\end{align}

By (\ref{17.22}),
\begin{eqnarray}
J_W e^{itK_W}=e^{itK_W} J_W. \label{20}
\end{eqnarray}
But $J_W$ is anti-unitary. Hence,
\begin{eqnarray}
J_W K_W=-K_W J_W. \label{21}
\end{eqnarray}

We now come to the anti-linear involutions $S_W$ and $S_{W^\prime}=J_WS_WJ^{-1}_W=J_WS_WJ_W$. They pick out the real
subspaces $\mathcal{H}_{W,W^\prime}$ and the associated Weyl algebras $\mathcal{W}(W)$, $\mathcal{W}(W^\prime)$. They commute,
as required by causality, and as shown below.

Consider
\begin{eqnarray}
e^{-\pi K_W}=\Delta_W^{1/2}. \label{19.1}
\end{eqnarray}
This operator is defined by the analytic continuation of $e^{itK_W}$ to the strip
\begin{eqnarray}
0<\textrm{Im} \, t<\pi. \label{19.2}
\end{eqnarray}
We will see that this continuation is possible.

The operator $S_W$, the Tomita-Takesaki operator, is given by
\begin{eqnarray}
S_W=J_W\Delta_W^{1/2}. \label{19.3}
\end{eqnarray}
Since
\begin{eqnarray}
J_W\Delta_W^{1/2}=\Delta_W^{-1/2}J_W \label{19.4}
\end{eqnarray}
by (\ref{21}), 
\begin{eqnarray}
S_WS_W=J_W\Delta_W^{1/2}J_W\Delta_W^{1/2}=J_W^2=\mathbb{I}, \label{19.5}
\end{eqnarray}
so that $S_W$ is an anti-linear involution. But it is {\it not} anti-unitary, since
\begin{eqnarray}
S_W^\dagger S_W=\Delta_W. \label{19.6}
\end{eqnarray}
$\Delta_W$ is self-adjoint, but not unitary. The operator $K_W$ has neither upper nor lower bound.
Hence, $\Delta_W$ is not bounded above, just like the Hamiltonian.

The real subspace $\textrm{Re}\,\mathcal{H}$ for $W$, which we denote by $(\textrm{Re}\,\mathcal{H}(W):=
\textrm{Re}\,\mathcal{H}(W)$, is determined by $S_W$:
\begin{eqnarray}
\zeta_W\in \textrm{Re}\,\mathcal{H}(W) \quad \iff \quad S_W\zeta_W=\zeta_W. \label{20.1}
\end{eqnarray}

As for $W^\prime$, by covariance,
\begin{eqnarray}
S_{W^\prime}=J_WS_WJ_W=\Delta_W^{1/2}J_W=J_W\Delta_W^{-1/2}, \label{20.2}
\end{eqnarray}
so that 
\begin{eqnarray}
\eta_{W^\prime}\in \textrm{Re}\,\mathcal{H}(W^\prime) \quad \iff \quad S_{W^\prime}\eta_{W^\prime}=\eta_{W^\prime}.
\end{eqnarray}

We now come to the crucial result.

\subsection{Causality}

This requires the proof that the Weyl algebras $\mathcal{W}_{W,W^\prime}$ are commutants of each other:
\begin{eqnarray}
\mathcal{W}_{W^\prime}=\mathcal{W}^\prime{}_W, \label{21.1}
\end{eqnarray}
the superscript prime denoting commutant.
There is also a change of notation: the spacetime region labeling the Weyl algebra is being put as a subscript.

Here, we will only prove that
\begin{eqnarray}
\mathcal{W}(\eta_{W^\prime})\mathcal{W}(\zeta_W)=\mathcal{W}(\zeta_W)\mathcal{W}(\eta_{W^\prime}). \label{21.2}
\end{eqnarray}
Since
\begin{eqnarray}
\mathcal{W}(\eta_{W^\prime})\mathcal{W}(\zeta_W)=\mathcal{W}(\eta_{W^\prime}+\zeta_W)e^{\frac{i}{2}
\textrm{Im}(\eta_{W^\prime}\zeta_W)}, \label{21.3}
\end{eqnarray}
we must verify that 
\begin{eqnarray}
(\eta_{W^\prime},\zeta_W)\in\mathbb{R}. \label{21.4}
\end{eqnarray}
For this purpose, we need the identity
\begin{eqnarray}
(J_W\alpha, J_W\beta)=(\beta,\alpha), \quad\quad \alpha,\beta\in\mathcal{H}, \label{21.5}
\end{eqnarray}
since $J_W$ is anti-unitary.

Now
\begin{eqnarray}
(\eta_{W^\prime},\zeta_W)=(S_{W^\prime}\eta_{W^{\prime}},S_W\zeta_W)
=(J_W\Delta_W^{-1/2}\eta_{W^\prime},J_W\Delta_W^{1/2}\zeta_W).
\end{eqnarray}
By (\ref{19.3}) and (\ref{20.2}),
\begin{eqnarray}
(\eta_{W^\prime},\zeta_W)=(\Delta_W^{1/2}\zeta_W,\Delta_W^{-1/2}\eta_{W^\prime}).
\end{eqnarray}
By (\ref{21.5}),
\begin{eqnarray}
(\eta_{W^\prime},\zeta_W)=(\zeta_W,\eta_{W^\prime}), \label{22.1}
\end{eqnarray}
since $\Delta_W^{\pm1/2}$ are self-adjoint.

Hence, $(\eta_{W^\prime},\zeta_W)$ is real and causality is established.

It is important to note that the causal complement $\mathcal{W}_{W^\prime}$ of $\mathcal{W}_W$ is its ``symplectic
complement''. Also, nowhere have we tried to localise states.

\subsection{On the Tomita-Takesaki Operator}

Let us denote the representation of the Weyl algebra $\mathcal{W}_W$ by the same symbol.

The Fock space representation of $\mathcal{W}_W$ is built from the vacuum state $e_0$. It has the following important
properties: it is {\it cyclic} and {\it separating}.

``Cyclic'' means that the Weyl algebra (complex linear combinations of all $\mathcal{W}(h)$) acting on $e_0$ gives the full
Hilbert space $\mathcal{H}$ on closure. 

``Separating'' means that if $\alpha\in\mathcal{W}_W$ annihilates $e_0$, then $\alpha=0$:
\begin{eqnarray}
\alpha e_0=0 \quad \iff \quad \alpha=0.
\end{eqnarray}

The implications of the remarkable results of Tomita-Takesaki theory are as follows. Since $e_0$ is cyclic and
separating for the $(\mathcal{C}^\ast-)$ algebra $\mathcal{W}_W$, there exists a unique anti-linear involution
$\tilde{S}_W$, 
\begin{eqnarray}
\tilde{S}_W^2=\mathbb{I}
\end{eqnarray}
with the property
\begin{eqnarray}
\tilde{S}_W\alpha e_0=\alpha^\ast e_0. \label{23.1}
\end{eqnarray}

Our $S_W$ fulfills (\ref{23.1}) since by (\ref{11.2}),
\begin{eqnarray}
S_W\mathcal{W}(h)e_0=S_W e^{-\|h\|^2/4}e^{ih/\sqrt{2}}e_0= e^{-\|h\|^2/4}e^{ih/4}e_0=\mathcal{W}^\ast(h)e_0, \label{23.2}
\end{eqnarray}
where we used $S_We_0=e_0$. 

Hence,
\begin{eqnarray}
\tilde{S}_W=S_W. \label{24.1}
\end{eqnarray}

The polar decomposition of $S_W$ is just
\begin{eqnarray}
S_W=J_W\Delta_W^{1/2}. \label{24.2}
\end{eqnarray}

The unitary group
\begin{eqnarray}
U(t)=\Delta_W^{it}, \label{24.3}
\end{eqnarray}
which leaves the vacuum invariant, generates the ``modular automorphism'' of $\mathcal{W}_W$. Since
\begin{eqnarray}
\Delta_W^{it}=e^{itK_W}, \label{24.33}
\end{eqnarray}
where $W$ is a wedge in this case, the boost group is the modular automorphism group and $K_W$ can be called
the modular Hamiltonian.

But below, when we sharpen the localisation from wedges to smaller regions $\mathcal{O}$, the corresponding operators
$S_{\mathcal{O}}=J_{\mathcal{O}}\Delta_{\mathcal{O}}^{1/2}$ with $S_{\mathcal{O}}^2=\mathbb{I}$ and
$\Delta_{\mathcal{O}}^{it}$ all exist, but in general do not have geometric interpretation.
 
\subsection{Remarks on the Real Subspaces of $S_{\mathcal{O}}$}

As before, we can define the real subspace $\textrm{Re}\, \mathcal{H}(\mathcal{O})$ of $\mathcal{H}$ using
$S_{\mathcal{O}}$. It is also standard:
\begin{eqnarray}
\overline{\textrm{Re}\, \mathcal{H}(\mathcal{O})+i\textrm{Re}\, \mathcal{H}(\mathcal{O})}=\mathcal{H}. \label{26}
\end{eqnarray}
The converse is also true: if $S_{\mathcal{O}}$ leads to a standard real subspace $\textrm{Re}\, \mathcal{H}(\mathcal{O})$
of $\mathcal{H}$, it fulfills (\ref{23.2}).

\section{Operators localised in $W$}

We need a simple definition of operators $\mathcal{W}_W(h)$ when $h\in\textrm{Re}\, \mathcal{H}(W)$. We can obtain it
by first recalling an elementary result in Fourier transforms.

Consider the Fourier transform $f$ of a function $\tilde{f}$ of $x$ which is supported on the half-line:
\begin{eqnarray}
f(\omega)=\int_0^\infty dx\, \tilde{f}(x)\, e^{i\omega x}. \label{26.1}
\end{eqnarray}
This integral converges if $\omega$ is continued into a complex variable with $\textrm{Im}\,\omega>0$. It is holomorphic
if $\textrm{Im}\,\omega>0$.

The elements $\alpha$ of the real subspace $\textrm{Re}\,\mathcal{H}(W)$ can be constructed in a similar manner.
We can find them by starting with
\begin{eqnarray}
\alpha_W(p)=\int_0^\infty dx_+\,dx_-\, \tilde{\alpha}_W(x_+,x_-)\, e^{i(p^0x_0+p^1x_1)}, \quad\quad x_\pm=x_1\pm x_0
\label{26.2}
\end{eqnarray}
where we have suppressed the variables $x_r$ $(r=2,3)$. In $W$, $x_\pm\geq0$ so that the integral is over $W$.
The representation $\rho$ of $\mathcal{P}_+$ can clearly be realised using the complex function $\alpha_W$ of momentum $p$.
We now argue that for positive energy representations,
\begin{eqnarray}
p_0\geq p_1>0, \label{27.1}
\end{eqnarray}
 $S_W$ can be applied on  $\alpha_W$.
The requirement
\begin{eqnarray}
S_W\alpha_W=\alpha_W \label{27.2}
\end{eqnarray}
then implies that
\begin{eqnarray}
\tilde{\alpha}_W(x_+,x_-)\in\mathbb{R}. \label{27.3}
\end{eqnarray}

The real subspace $\textrm{Re}\,\mathcal{H}(W)$ is thus spanned by functions $\alpha_W$ with real Fourier transforms
which are supported in $W$.

Let us show this result. With $p_\pm=p^0\pm p^1>0$, as is the case in positive energy representations,
\begin{eqnarray}
\alpha_W(p)=\int_0^\infty dx_+\, dx_-\, \tilde{\alpha}_W(x_+,x_-)\, e^{i(p_+x_+-p_-x_-)/2}. \label{27.4}
\end{eqnarray}

Under the boost transformation
\begin{eqnarray}
\Lambda_W(t)=\left(
\begin{array}{ccccc}
\cosh t & -\sinh t & 0 & 0 \\
-\sinh t & \cosh t & 0 & 0 \\
0 & 0 & 1 & 0 \\
0 & 0 & 0 & 1
\end{array}
\right), \label{28.1}
\end{eqnarray}
we have
\begin{eqnarray}
\rho(\Lambda_W(t))=e^{itK_W} \label{28.2}
\end{eqnarray}
and
\begin{eqnarray}
(e^{itK_W}\alpha_W)(p)= \alpha_W(\Lambda^{-1}_W(t)p)=\int_0^\infty dx_+\,dx_-\, \tilde{\alpha}_W(x_+,x_-)\,
e^{i(e^t p_+x_+-e^{-t}p_-x_-)}. \label{28.3}
\end{eqnarray}
For
\begin{eqnarray}
t=i\mu, \quad\quad 0<\mu<\pi, \label{28.4}
\end{eqnarray}
we get
\begin{eqnarray}
(e^{-\mu K_W}\alpha_W)(p)=\int_0^\infty dx_+\, dx_-\, \tilde{\alpha}_W(x_+,x_-)\,
e^{i\cos\mu(p_+x_+-p_-x_-)}\, e^{-\sin\mu(p_+x_++p_-x_-)}. \label{28.5}
\end{eqnarray}
The first exponential has modulus 1, while the second is a damping factor in the interval (\ref{28.4}), since
$p_\pm,x_\pm>0$. Thus, (\ref{28.3}) is the boundary value $\mu\downarrow0$ of a holomorphic function in the strip
\begin{eqnarray}
0<\textrm{Im}t<\pi. \label{29.6}
\end{eqnarray}
When 
\begin{eqnarray}
\textrm{Im}t\uparrow\pi, \label{29.7}
\end{eqnarray}
we get
\begin{eqnarray}
(e^{-\pi K_W}\alpha_W)(p)=(\Delta_W^{1/2}\alpha_W)(p)=\int_0^\infty dx_+\, dx_-\, \tilde{\alpha}_W(x_+,x_-)\,
e^{-i(p_+x_++p_-x_-)}.
\label{29.8}
\end{eqnarray}
Hence,
\begin{eqnarray}
(S_W\alpha_W)(p)= (J_W\Delta_W^{1/2}\alpha)(p)=\int_0^\infty dx_+\, dx_-\, \overline{\tilde{\alpha}}_W(x_+,x_-)\,
e^{i(p_+x_++ip_-x_-)} \label{29.9}
\end{eqnarray}
and the condition $S_W\alpha_W=\alpha_W$ implies that $\tilde{\alpha}_W(x_+,x_-)\in\mathbb{R}$, as claimed.

\subsection{Remarks}

\begin{itemize}
\item The above analyticity and hence the existence of $S_W$ and localisation can be established only for ``positive
energy representations'', where  $(p_0-|\vec{p}|)\geq0$.
\item From (\ref{29.8}), $\Delta_W^{1/2}$ is seen to reverse the sign of energy.
In Feynman's language, it converts an outgoing particle line into an incoming anti-particle line in a scattering
diagram. We illustrate this interpretation  in figure 2. 
\begin{figure}[h]
\begin{center}
\includegraphics[scale=0.5]{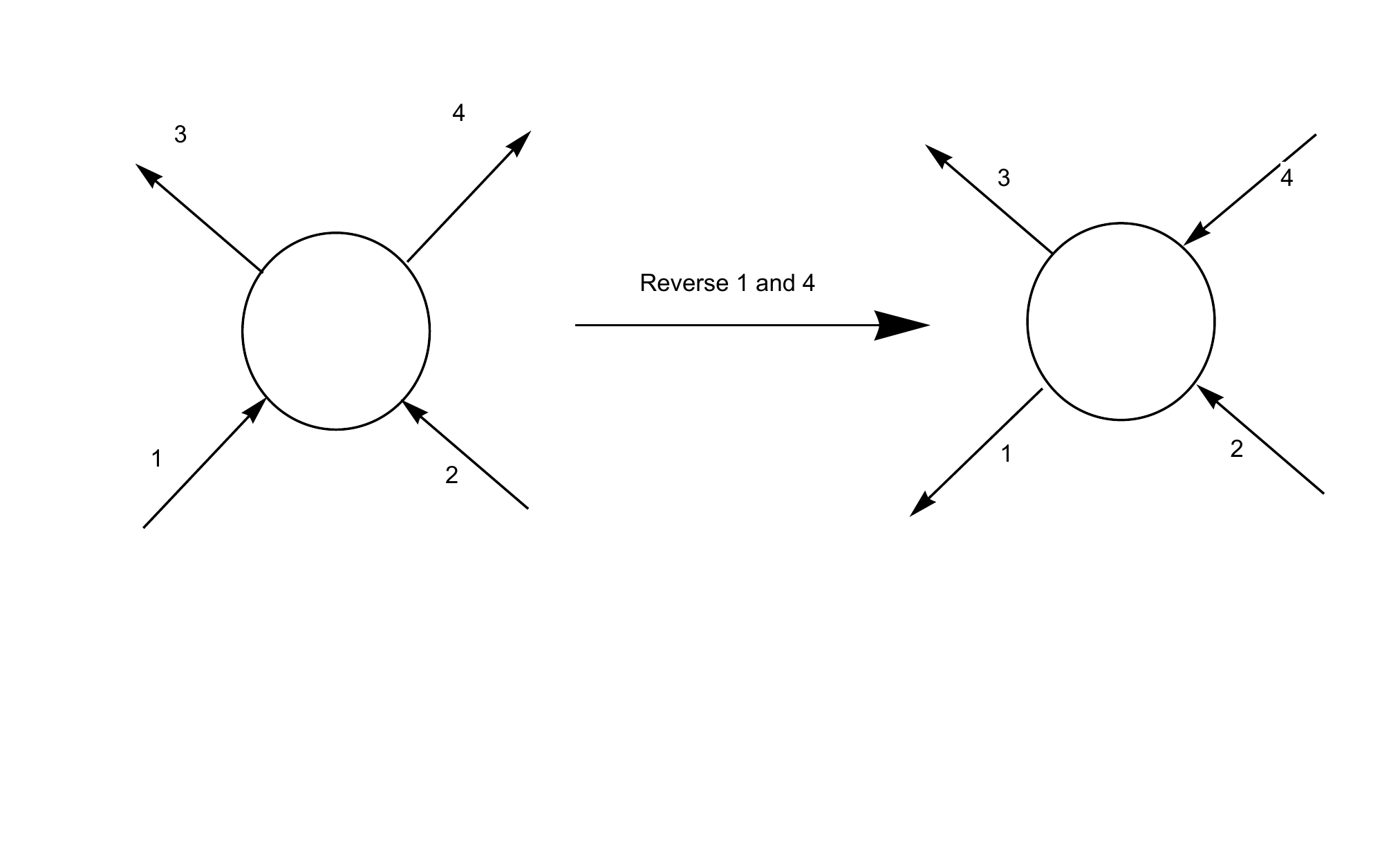}
\caption{$\Delta_W^{1/2}$ is seen to reverse the sign of energy}
\end{center}
\end{figure}

Thus, as Fassarella and Schroer \cite{SF} have emphasised, $\Delta_W^{1/2}$ seems related to crossing symmetry.
\end{itemize}

All this means in particular that localisation requires anti-particles (which may be the same as particles).

\section{Sharpening Localisation}

Wedge localisation is rather weak as a wedge is not even compact. One would like localisation in spacetime regions
$\mathcal{O}$ of arbitrary small size.

For this purpose, first consider the intersection of two wedges $W_1$ and $W_2$ producing the ``causal diamond'' (shown in figure 3). 
\begin{figure}[h]
\begin{center}
\includegraphics[scale=.7]{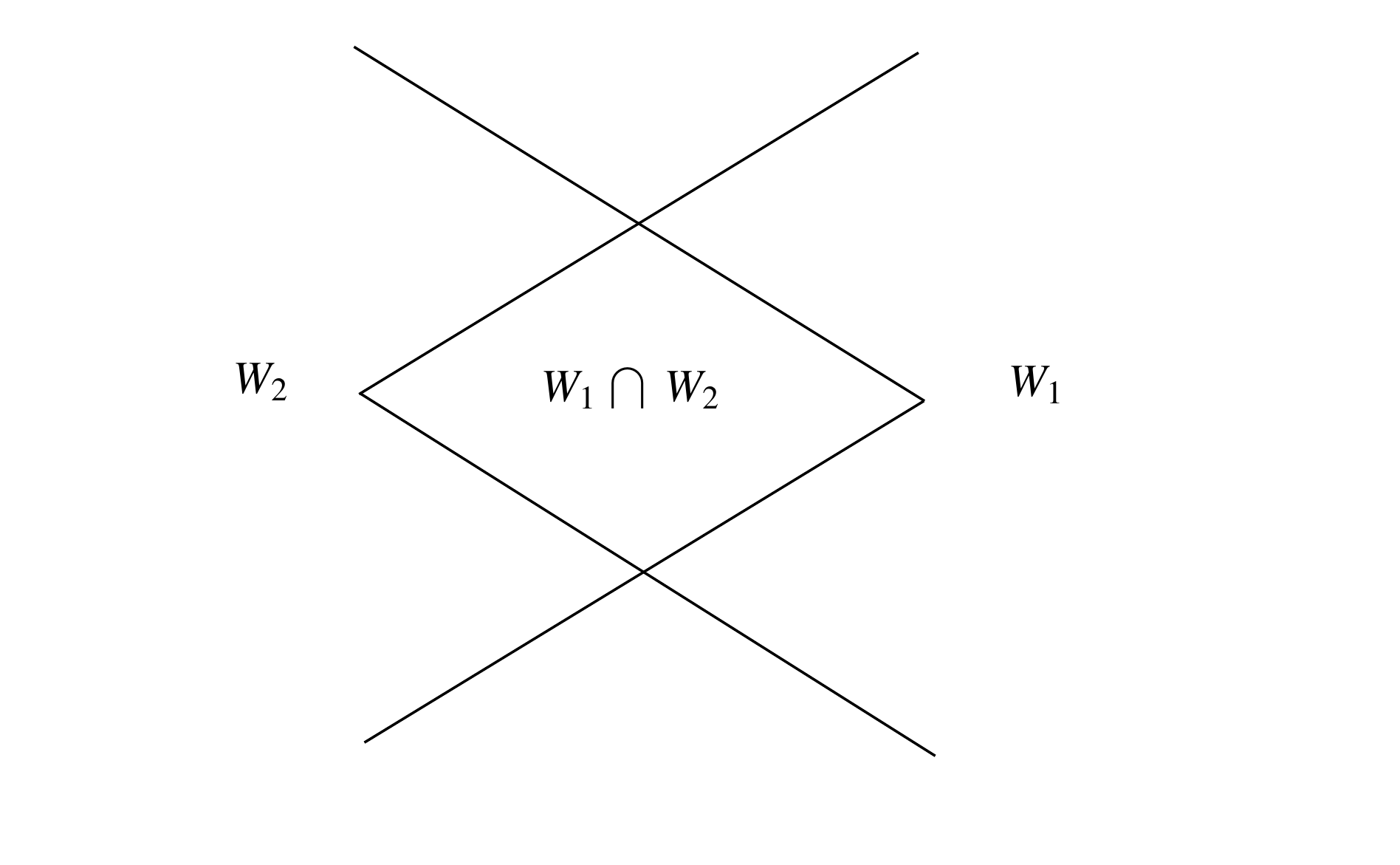}
\caption{Causal Diamond}
\end{center}
\end{figure}
We can then consider the associated real Hilbert space:
\begin{eqnarray}
\textrm{Re}\,\mathcal{H}(W_1\cap W_2):=\textrm{Re}\,\mathcal{H}(W_1)\cap \textrm{Re}\,\mathcal{H}(W_2).
\end{eqnarray}
One then shows that $\textrm{Re}\,\mathcal{H}(W_1\cap W_2)$ is standard:
\begin{eqnarray}
\begin{array}{llll}
&\mathcal{H}=\textrm{Re}\,\mathcal{H}(W_1\cap W_2)\oplus i\textrm{Re}\,\mathcal{H}(W_1\cap W_2), \\ \\
&\textrm{Re}\,\mathcal{H}(W_1\cap W_2) \cap i\textrm{Re}\,\mathcal{H}(W_1\cap W_2)=\{0\} \label{31.1}
\end{array}
\end{eqnarray}
(where taking closure is understood).

Now (\ref{31.1}) is enough to define the modular operator and show causality.

Thus, if $\zeta\in\mathcal{H}$, we have the unique decomposition
\begin{eqnarray}
\zeta=\textrm{Re}\,\zeta_{W_1\cap W_2}+i\textrm{Im}\, \zeta_{W_1\cap W_2}, \label{22.11}
\end{eqnarray}
where the first term is in $\textrm{Re}\, \mathcal{H}_{W_1\cap W_2}$ and the second in $i\textrm{Re}\,
\mathcal{H}_{W_1\cap W_2}$.

The modular involution $S_{W_1\cap W_2}$ is then defined by
\begin{eqnarray}
S_{W_1\cap W_2}\zeta=\textrm{Re}\, \zeta_{W_1\cap W_2}-i\textrm{Im}\,\zeta_{W_1\cap W_2}. \label{22.2}
\end{eqnarray}
The definition of polar decomposition of $S_{W_1\cap W_2}$,
\begin{eqnarray}
S_{W_1\cap W_2}=J_{W_1\cap W_2}\Delta_{W_1\cap W_2}^{1/2}, \label{22.3}
\end{eqnarray}
shows that
\begin{eqnarray}
\Delta_W= S_{W_1\cap W_2}^\dagger S_{W_1\cap W_2}, \label{23.4}
\end{eqnarray}
where the RHS can be calculated from (\ref{22.2}). Then, from (\ref{22.3}), we have $J_{W_1\cap W_2}$.

Just as before as in the case of $S_{W^\prime}$, one shows that the modular involution
\begin{eqnarray}
J_{W_1\cap W_2} S_{W_1\cap W_2} J_{W_1\cap W_2}=\Delta_{W_1\cap W_2}^{1/2} J_{W_1\cap W_2} \label{33.1}
\end{eqnarray}
determines the causal complement $\mathcal{W}^\prime_{W_1\cap W_2}$ of 
$\mathcal{W}_{W_1\cap W_2}$.

In this way, we have the algebra $\mathcal{W}_{W_1\cap W_2}$ localised in $W_1\cap W_2$.

We can even characterise the elements $\alpha_{W_1\cap W_2}$ in $\textrm{Re}\,\mathcal{H}_{W_1\cap W_2}$: we use (\ref{26.2}),
but with real functions $\tilde{\alpha}_{W_1\cap W_2}$ supported in $W_1\cap W_2$.


\subsection{Further Sharpening of Localisation}

A spacetime region $\mathcal{O}$ is said to be {\it causally complete} if the following condition is satisfied:
Let $\mathcal{O}^\prime$ denote the causal complement of $\mathcal{O}$ so that points of $\mathcal{O}^\prime$
are spacelike separated from $\mathcal{O}$.  Let $\mathcal{O}^{\prime\prime}$ be the causal complement of
$\mathcal{O}^\prime$. Then, $\mathcal{O}$ is causally complete if $\mathcal{O}^{\prime\prime}=\mathcal{O}$.
The diamond in the last figure above is causally complete. 

A causally complete region $\mathcal{O}$ is known to be the intersection of wedges. As wedges are mutually related by the action of
$\mathcal{P}_+$, causally complete regions are invariant under the action of $\mathcal{P}_+$. They form a covariant net. 

Given this net, we can localise the Weyl algebra to $\mathcal{O}$, to obtain $\mathcal{W}_{\mathcal{O}}$.

We now show how to explicitly construct $\mathcal{W}_{\mathcal{O}}$. It involves the construction of the standard
real subspace $\textrm{Re}\,\mathcal{H}_{\mathcal{O}}\subset\mathcal{H}$.

We can describe the elements of $\textrm{Re}\,\mathcal{H}_{\mathcal{O}}$ using (\ref{26.2}), but with the real functions
$\tilde{\alpha}_{\mathcal{O}}$ now supported in $\mathcal{O}$.

The Weyl algebra is then constructed from its elements as described earlier.

It is important to know also the modular involution $S_{\mathcal{O}}$ and the causal
complement of $\mathcal{W}_{\mathcal{O}}$.

As for $S_{\mathcal{O}}$, we follow (\ref{22.11})-(\ref{22.3}), but with $W_1\cap W_2$ replaced by $\mathcal{O}$.
That gives us $J_{\mathcal{O}}$ and $\Delta_{\mathcal{O}}^{1/2}$ in $S_{\mathcal{O}}=J_W\Delta_{\mathcal{O}}^{1/2}$.

The causal complement $\mathcal{W}^\prime_{\mathcal{O}}$ of $\mathcal{W}_{\mathcal{O}}$ is then
\begin{eqnarray}
J_{\mathcal{O}}\mathcal{W}_{\mathcal{O}}J^{-1}_{\mathcal{O}}=J_{\mathcal{O}}\mathcal{W}_{\mathcal{O}}J_{\mathcal{O}}.
\label{35.1}
\end{eqnarray}

\subsection{Remarks}
We can show that the vacuum state $|0> :=|e_0>$ restricted to the observables in the Rindler wedge $W$ is mixed : it is a thermal or KMS state. This is Unruh's result.The proof is as follows.

First recall the KMS condition. In terms of a density matrix $\rho = exp(-\beta H)$, it is
\begin{eqnarray}
\omega_\rho(AB) := Tr ( \rho AB)/(Tr \rho ) = \omega_\rho(B exp(-\beta H)Aexp(\beta H))=\omega_\rho (B U_{i\beta}(A))
\end{eqnarray}
where   $U_{i\beta}(A) =A$ evolved for  imaginary time $i\beta$.

A state  $\omega_\beta$ is KMS if it fulfills 
\begin{eqnarray}
\omega_\beta(AB)= \omega_\beta(B U_{i\beta}(A))
\end{eqnarray}
even if this state does not come from a density matrix .

 It is not difficult to show that
\begin{eqnarray}
S_W V(h) |e_0> =V(h)^*|e_0>.
\end{eqnarray}
Hence 
\begin{eqnarray}
 \hspace{-1cm}<e_0| AB|0> &=& <A^*e_0 |Be_0> =<J_W \sqrt{\Delta_W} Ae_0|J_W\sqrt{\Delta_W} B^*e_0 >\\
  &=&<\sqrt{\Delta_W} B^*e_0 |\sqrt{\Delta_W} Ae_0>  \,\,\,  \,\,\,( \textrm{as } J_W \textrm{ is anti-unitary}) \\ 
  &=& <e_0|B{\Delta_W}A |e_0> \,\,\,  \,\,\,(\textrm{as }\sqrt{ \Delta_W} \textrm{ is self-adjoint}) \\
  &  =&<e_0|B\Delta_WA \Delta_W^{-1} |e_0> \,\,\, \,\,\,  (\textrm{as }  |e_0>  \textrm{is invariant under }  \Delta_W^{-1}).
\end{eqnarray}
 Thus $|e_0><e_0|$ is a mixed KMS state for the algebra $\mathcal{W}_W$ and for the `Hamiltonian' $H= 2 \pi K_W/\beta$.

But since the spectrum of $K_W$ is unbounded above and below,  $\Delta_W$ is not of trace class and we cannot construct a density matrix like $\rho$ above for this state.
\begin{itemize}
\item It is known that 
\begin{eqnarray}
&& \textrm{Re}\,\mathcal{H}_{\mathcal{O}}=\bigcap_{W\supset\mathcal{O}}\textrm{Re}\,\mathcal{H}_W, \label{35.2} \\
&& \mathcal{W}_{\mathcal{O}}=\bigcap_{W\supset\mathcal{O}}\mathcal{W}_W. \label{35.3}
\end{eqnarray}
\item We can show as before that the vacuum defines a KMS state for the Hamiltonian $2\pi K_{\mathcal{O}}/\beta$,
where $\Delta_{\mathcal{O}}=e^{-2\pi K_{\mathcal{O}}}$.

But when $\mathcal{O}$ is not a wedge, not even when it is a causal diamond, $K_{\mathcal{O}}$ has no known geometrical
meaning. It is a boost generator of $\mathcal{P}_+$ only when $\mathcal{O}$ is a wedge.
\item The theory shows that $\Delta_{\mathcal{O}} e_0=e_0$ or that 
\begin{eqnarray}
e^{itK_{\mathcal{O}}}e_0=e_0 \label{36.1}
\end{eqnarray}
for every causally complete $\mathcal{O}$. Thus we get an infinite number of localised boost groups
$\Lambda_{\mathcal{O}}=\{ e^{itK_{\mathcal{O}}}\}$ labelled by the causally complete net, all of which leave the vacuum
invariant, just like $\Lambda_W$.
Their localisation reminds us of gauge groups, but the latter either act trivially  on {\it all} quantum states or define
superselection sectors. Neither is the case with $\Lambda_{\mathcal{O}}$.

The physical meaning of $\Lambda_{\mathcal{O}}$'s has not been understood.
\end{itemize}

\section{Introducing Spin}

In these notes, we have not treated the construction of the UIRR's of $\mathcal{P}_+$ using Wigner's approach.
For this reason, we will treat only the spin $1/2$ case, assumming familiarity with the construction of its UIRR.
We refer to~\cite{BalBoo} for example for further details.

For relativistic particles with spin, the transformation properties of a state vector with definite momentum involves 
the Wigner boost and Wigner rotation. Their presence spoils the analyticity property of $e^{itK_W}$ in the strip
$0<\textrm{Im} t<\pi$. Localisation for such representations involves additional considerations.

We illustrate the situation for a UIRR of $\mathcal{P}_+$ with spin $1/2$ and mass $m>0$.

\subsection{Massive Particle of Spin $1/2$}

\subsubsection{Preliminaries}

Let
\begin{eqnarray}
\hat{p}=(m,\vec{0}) \label{37.1}
\end{eqnarray}
be the standard momentum. A basic ingredient in setting up the UIRR is the choice of the Wigner boost
$L_p\in\mathcal{P}_+^\uparrow$ which transforms $\hat{p}$ to momentum $p$ (see~\cite{BalBoo}):
\begin{eqnarray}
L_p\hat{p}=p. \label{37.2}
\end{eqnarray}

A convenient choice of $L_p$ uses the $2\times2$ representation of $p$:
\begin{eqnarray}
p\rightarrow \sigma\cdot p=(\sigma\cdot p)^\dagger,
\quad\quad \sigma=(\sigma^0=\mathbb{I}, \sigma^i=\textrm{Pauli matrices}). \label{38.1}
\end{eqnarray}
In this representation, $\mathcal{P}_+^\uparrow$ acts by elements $g\in SL(2,\mathbb{C})$:
\begin{eqnarray}
\sigma\cdot p\rightarrow g\sigma\cdot p g^\dagger. \label{38.2}
\end{eqnarray}
For rotations, $g\in SU(2)$, so that $g^\dagger =g^{-1}$. For boosts, $g=g^\dagger$. Thus, since
\begin{eqnarray}
\sigma\cdot \hat{p}=m\mathbb{I} \label{38.3}
\end{eqnarray}
and
\begin{eqnarray}
\sigma\cdot p >0, \label{38.4}
\end{eqnarray}
that is, its eigenvalues are positive, as may be verified, we can choose for the boosts,
\begin{eqnarray}
g(p)=\left(\frac{\sigma\cdot p}{m}\right)^{1/2}\in SL(2,\mathbb{C}), \label{38.5}
\end{eqnarray}
where the square root is the positive one. Thus,
\begin{eqnarray}
\sigma\cdot (L_p\hat{p})=g(p)(\sigma\cdot \hat{p})g^\dagger (p). \label{38.55}
\end{eqnarray}

The transformations in $\mathcal{P}_+^\uparrow$ leaving $\hat{p}$ invariant is $SO(3)$. In the $2\times2$
$SL(2,\mathbb{C})$ representation, it becomes the UIRR $D^{1/2}$ of $SU(2)$ for angular momentum $1/2$.
In the Wigner approach, we first introduce the vectors $|\hat{p},\lambda\rangle$. If the UIRR is $U$
and $h\in SU(2)$, we set
\begin{eqnarray}
U(h)|\hat{p},\lambda\rangle = |\hat{p},\rho\rangle D_{\rho\lambda}^{1/2}(h), \quad\quad
U(L_p)|\hat{p},\lambda\rangle =|p,\lambda\rangle. \label{39.1}
\end{eqnarray}
Using the $2\times 2$ matrix rotation, we can change notation as follows:
\begin{eqnarray}
\begin{array}{llll}
&&|\hat{p},\lambda\rangle\rightarrow|\sigma\cdot \hat{p},\lambda\rangle, \\ \\
&&U(h)|\sigma\cdot \hat{p},\lambda\rangle = |\sigma\cdot\hat{p},\rho\rangle D_{\rho\lambda}^{1/2}(h), \\ \\
&&U(L_p)|\sigma\cdot\hat{p},\lambda\rangle =|\left(\frac{\sigma\cdot p}{m}\right)^{1/2}\sigma\cdot\hat{p}
\left(\frac{\sigma\cdot p}{m}\right)^{1/2},\lambda\rangle=|\sigma\cdot p,\lambda\rangle. \label{39.2}
\end{array}
\end{eqnarray}

It is a consequence of (\ref{39.2}) that if $g\in SL(2,\mathbb{C})$,
\begin{eqnarray}
U(g)|\sigma\cdot p,\lambda\rangle=|\sigma\cdot\Lambda(g)p,\rho\rangle D_{\rho\lambda}^{1/2}(h(p,g)), \label{39.3}
\end{eqnarray}
where $\Lambda(g)$ the Lorentz transformation associated with $g$,
\begin{eqnarray}
g\sigma\cdot p g^\dagger=\sigma\cdot \Lambda(g)p
\end{eqnarray}
and $h(p,g)\in SU(2)$ is called the Wigner rotation:
\begin{eqnarray}
h(p,g)=\left(\frac{\sigma\cdot \Lambda(g)p}{m}\right)^{-1/2}g\left(\frac{\sigma\cdot p}{m}\right)^{1/2}. \label{40.1}
\end{eqnarray}

For further details, see~\cite{BalBoo}.

Thus, if $g$ is the boost $e^{-itK_W}$, which becomes $e^{t\sigma_1/2}$ in the $2\times2$ $SL(2,\mathbb{C})$ representation,
\begin{eqnarray}
h(p,e^{t\sigma_1/2})=\left(\frac{\sigma\cdot e^{itK_W}p}{m}\right)^{-1/2}e^{t\sigma_1/2}
\left(\frac{\sigma\cdot p}{m}\right)^{1/2}.
\label{40.2}
\end{eqnarray}

\subsubsection{Analyticity}

We need the analyticity of (\ref{40.2}) in the strip $0<\textrm{Im}t<\pi$ ( See (6.12)).
But for $p_r(r\geq2)\neq0$, this requirement is not met, leading to an obstruction to localisation.

The way around it is as follows. Let us imbed the $D^{1/2}$ UIRR of $SU(2)$ in the $D^{(1/2,0)}$ UIRR of $SL(2,\mathbb{C})$.
Then, we can write
\begin{eqnarray}
D^{(1/2,0)}(h(p,g))=D^{(1/2,0)}\left[\left(\frac{\sigma\cdot\Lambda(g)p}{m}\right)^{1/2}\right]^{-1} D^{(1/2,0)}(g)
D^{(1/2,0)}\left(\frac{\sigma\cdot p}{m}\right)^{1/2}. \label{41.1}
\end{eqnarray}
Using this decomposition, let us define
\begin{eqnarray}
|\sigma\cdot p,\lambda\rangle^{\ast}=|\sigma\cdot p,\rho\rangle D^{(1/2,0)}\left[\left(
\frac{\sigma\cdot p}{m}\right)^{1/2}\right]^{-1}. \label{41.2}
\end{eqnarray}
It follows from (\ref{39.3}) and (\ref{41.1}) that
\begin{eqnarray}
U(g)|\sigma\cdot p,\lambda\rangle^{\ast}=|\sigma\cdot \Lambda(g)p,\rho\rangle^{\ast}
D^{(1/2,0)}_{\rho\lambda}(g). \label{41.3}
\end{eqnarray}

Thus, by working with functions $f_\lambda$ of $p$ (or $\sigma\cdot p$) with the transformation (\ref{41.3}),
we can remove the obstruction to analyticity encountered above.

\subsubsection{Causality}

For the spin 0 case we treated above, the quantum field which emerges {\it commutes} for spacelike separations.
We can see this as follows.

In the notation of (\ref{26.2}), let
\begin{eqnarray}
\varphi(\tilde{\alpha}_W)=a^\ast (\alpha_W)+a(S_W\alpha_W), \label{42.1}
\end{eqnarray}
where $\tilde{\alpha}_W$ is supported in $W$ and is real, as before. So $S_W\alpha_W=\alpha_W$, but we put in
$S_W$ for later convenience.
We also set
\begin{eqnarray}
[a(S_W\alpha_W),a^\ast(\beta_W)]=(S_W\alpha_W,\beta_W), \label{42.3}
\end{eqnarray}
where $\tilde{\beta}_W$ also has support $W$. All other commutators involving $a$ and $a^\ast$ vanish as usual.

Similarly,
\begin{eqnarray}
\varphi(\tilde{\alpha}_{W^\prime})=a^\ast (\alpha_{W^\prime})+a(S_{W^\prime}\alpha_{W^\prime}), \label{43.1}
\end{eqnarray}
where
\begin{eqnarray}
S_{W^\prime}=J_W S_W J_W=\Delta_W^{1/2} J_W. \label{43.2}
\end{eqnarray}
Hence,
\begin{eqnarray}
\nonumber
[\varphi(\tilde{\alpha}_W),\varphi^\ast (\tilde{\beta}_{W^\prime})]&=&
(S_W\tilde{\alpha}_W,\tilde{\beta}_{w^\prime})-(S_{w^\prime}\tilde{\beta}_{W^\prime},\tilde{\alpha}_W) \\
\nonumber
&=&(S_W\tilde{\alpha}_W,\tilde{\beta}_{W^\prime})-(\Delta_W^{1/2} J_W\tilde{\beta}_{W^\prime},\tilde{\alpha}_W)\\
&=&0,
\label{43.3}
\end{eqnarray}
where the anti-unitarity of $J_W$ has been used.

Thus, spacelike separated $\varphi$'s commute.

We now extend this analysis to spin $1/2$.
Fields of spin $1/2$ must {\it anti-commute} for spacelike separation, whereas the modular involution $S_W$ leads to a
commutation relation. Therefore, in the definition of a spin $1/2$ field $\psi$, we change $S_W$ to
\begin{eqnarray}
\mathcal{S}_W=iS_W. \label{43.4}
\end{eqnarray}

It too has the property
\begin{eqnarray}
\mathcal{S}_W^2=\mathbb{I}. \label{44.1}
\end{eqnarray}

Then, for a spin $1/2$ field $\psi$,
\begin{eqnarray}
\psi(\tilde{\alpha}_W):=a^\ast(\alpha_W)+a(\mathcal{S}_W\alpha_W), \label{44.2}
\end{eqnarray}
where we set
\begin{eqnarray}
[a(\mathcal{S}_W\alpha_W),a^\ast(\beta_W)]_+=(\mathcal{S}_W\alpha_W,\beta_W), \label{44.3}
\end{eqnarray}
with zero for the other anti-commutators.

For $W^\prime$, by covariance,
\begin{eqnarray}
\mathcal{S}_{W^\prime}=J_W\mathcal{S}_W J_W=-i J_W(J_W\Delta_W^{1/2})J_W=-i\Delta_W^{1/2} J_W. \label{44.4}
\end{eqnarray}
Therefore,
\begin{eqnarray}
\nonumber
[\psi(\tilde{\alpha}_W),\psi(\tilde{\beta}_{W^\prime})]_+&=&(\mathcal{S}_W\alpha_W,\beta_{W^\prime})
+(\mathcal{S}_{W^\prime}\beta_{W^\prime},\alpha_W) \\
\nonumber
&=&-i(J_W\Delta_W^{1/2}\alpha_W,\beta_{W^\prime})+i(\Delta_W^{1/2}J_W\beta_{W^\prime},\alpha_W) \\
&=&0, \label{44.5}
\end{eqnarray}
where the last line follows from anti-linearity of $J_W$.

The $i$ is the ``statistical'' factor which corrects the commutator to anti-commutator. Its square being $-1$, which
corresponds to
$2\pi$ rotation being $-1$, it accounts for the spin-statistics theorem.

\subsection{Final Remarks}

The Poincar\'{e} group has two ``exceptional'' classes of positive energy UIRR's.

One occurs in $3+1$ dimensions for massless particles where the little or stability group in general is
$\overline{E}(2)$,
the two-fold covering group of the Euclidean group. For particles like photons with two helicities, the translation part of
$\overline{E}(2)$ is represented trivially, by identity operators.

But there are UIRR's where  the translations of $\overline{E}(2)$ are represented non-trivially. 
In these UIRR's, helicity takes on all half-integral values for fermions and all integral values for bosons.
Particles characterised by such UIRR's are said to have continuous spin. 

The second class of ``exceptional'' UIRR's occurs in $2+1$ dimensional spacetime. They are the anyons.
For anyons, $2\pi$-rotation is neither $(+1)$ nor $(-1)$. Further, they obey braid statistics.
The latter is based on the braid group~\cite{unknown2} and not on the permutation group. Such particles,
which can occur as excitations in two-dimensional lattices of spins, are thought to be important for
``topological quantum computations''. 

If we exclude these exceptional UIRR's, for all other UIRR's of the Poincar\'{e} group, localisation in the manner
we have described works. Familiar local fields can also be constructed, as in \cite{SF,large}.

But that is not the case for the exceptional UIRR's~\cite{SF, large}. For such UIRR's, standard local fields, such as
$\varphi$ or $\psi$ above, do not exist. The best-localised fields are localised on ``strings''. Thus, such a field
$\chi$ for $2\pi$-rotation $+1$ say, is labelled by a spacetime position $x$ and a spacelike direction $e$:
\begin{eqnarray}
e\cdot e=-1. \label{47.1}
\end{eqnarray}
Both $x$ and $e$ transform under Lorentz transformations:
\begin{eqnarray}
\Lambda:\,\,\chi(x,e)\,\rightarrow\,\chi(\Lambda^{-1}x,\Lambda^{-1}e). \label{47.2}
\end{eqnarray}

As for causality, the condition is novel. Let
\begin{eqnarray}
x+\mathbb{R}^+e:= \{x+\lambda e: \,\, 0\leq\lambda<\infty\}. \label{47.3}
\end{eqnarray}
Thus, $x+\mathbb{R}^+e$ is a spacelike string from $x$ to $\infty$. Then, causality is expressed by
\begin{eqnarray}
[\chi_1(x,e),\chi_2(x^\prime,e^\prime)]=0
\end{eqnarray}
if $x+\mathbb{R}^+e$ is spacelike to $x^\prime+\mathbb{R}^+ e^\prime$, that is, each point $p$ of the former,
$p\in x+\mathbb{R}^+e$ is spacelike to each point $p^\prime$ of the latter, $p^\prime \in x^\prime +\mathbb{R}^+ e^\prime$.

The two-point function for such fields has been worked out.

Incidentally, such string-localised fields exist even for non-exceptional UIRR's~\cite{SF,large}. They have better
ultra-violet behaviour.

\subsection{Remarks}

\begin{itemize}
\item Dirac~\cite{Dirac} had long ago considered fields dependent on a spacelike direction in the  context of gauge theories.
Thus, for a $U(1)$ gauge theory with a charged field $\psi$ and electromagnetic connection $A$, he had defined the field
\begin{eqnarray}
\hat{\psi}(x,e)=\left[P\exp\left(i\int^x A_\mu(x') dx^{\prime\mu}\right)\right]\psi(x),
\end{eqnarray}
where the integral is along the line $x+\tau e$ as $\tau$ increases from $(-\infty)$ to 0.
The field $\hat{\psi}$ is invariant under the gauge transformation
\begin{eqnarray}
\psi(x)\rightarrow e^{i\Lambda(x)}\psi(x),
 \quad\quad
A_\mu(x+\tau e)\rightarrow A_\mu(x+\tau e)+(\partial_\mu\Lambda)(x+\tau e), \label{48.1}
\end{eqnarray}
with the usual condition $(\partial_\mu\Lambda)(x+\tau e)\rightarrow 0$ as $\tau\rightarrow-\infty$.

The field $\hat{\psi}$ of Dirac does not seem to be the string-localised field considered above. The latter is a free field
and not coupled to a gauge field.
\item It is a striking and important result that string-localised fields do not admit a Lagrangian description. They seem
to have no classical counterpart of a familiar sort.
\end{itemize}
\textbf{ Acknowledgements}

 I thank  Nirmalendu Acharyya and  Veronica Errasti Diez  for their invaluable help in the preparation of this manuscript. I also thank Nemani Suryanarayana and the Institute of Mathematical Sciences for hospitality while this work was being completed.

\end{document}